\documentclass[oneside,reqno,a4paper,11pt]{amsart}

\usepackage{color,soul,supertabular,longtable,verbatim}
\usepackage{amsmath,bbm,amsfonts,amssymb}
\usepackage{hyperref}
\usepackage{bibtopic}
\newtheorem {theorem}{Theorem}[section]
\newtheorem {lemma}[theorem]{Lemma}

\newtheorem {proposition}[theorem]{Proposition}
\theoremstyle{remark}

\theoremstyle{problem}

\theoremstyle{definition}

\theoremstyle{plain} \numberwithin {equation}{section}

\def\XXint #1#2#3{{\setbox 0=\hbox {$#1{#2#3}{\int }$}
\vcenter {\hbox {$#2#3$}}\kern -.5\wd 0}}

\def\ba{\begin{aligned}}
\def\ea{\end{aligned}}
\def\bn{\begin{enumerate}}
\def\en{\end{enumerate}}
\def\be{\begin{equation}}
\def\ee{\end{equation}}

\def\lap{\triangle}

\def\g{\gamma}
\def\a{\alpha}

\def\O{\Omega}

\def\ep{\epsilon}
\def\D{\mathcal{D}}

\def\p{\partial}

\def\norm[#1]#2{\|#2\|_{#1}}
\def\p{\partial}

\begin{document}




\vspace{1cm}

\title{A Blow-Up Criterion for the Compressible Navier-Stokes equations}
\author[Xiangdi Huang, Zhouping Xin]
{Xiangdi Huang, Zhouping Xin}

\address{ Xiangdi Huang \hfill\break\indent
          The Institute of Mathematical Sciences, The Chinese University of Hong Kong, Hong Kong}
          \email{\href{mailto:xdhuang@math.cuhk.edu.hk}{xdhuang@math.cuhk.edu.hk}}
\address{ Zhouping Xin \hfill\break\indent
          The Institute of Mathematical Sciences, The Chinese University of Hong Kong, Hong Kong}
          \email{\href{mailto:zpxin@ims.cuhk.edu.hk}{zpxin@ims.cuhk.edu.hk}}
\maketitle

\begin{abstract}
In this paper, we obtain a blow up criterion for strong solutions to
the 3-D compressible Naveri-Stokes equations just in terms of the gradient
of the velocity, similar to the Beal-Kato-Majda criterion
for the ideal incompressible flow. The key ingredients in our analysis are the a priori
super-norm estimate of the momentum by a Moser-iteration and an estimate of the
space-time square mean of the gradient of the density.
  In addition, initial vacuum is allowed in our case.
\end{abstract}


\section{introduction}

We shall study the following isentropic compressible $Navier-Stokes$
equations in $3-\D$ case:
\begin{equation}
\left\{ \ba
& \frac{\p\rho}{\p t} + \rm{div}(\rho u)=0,\\
& \frac{(\p\rho u)}{\p t} + \rm{div}(\rho u\otimes u)
-\mu\lap u-(\mu + \lambda)\nabla(\rm{div}u) + \nabla P(\rho)=0
\ea \right.
\end{equation}

Where $\rho, u, P$ denotes the density, velocity and pressure respectively.
The pressure-density state equation is given by
\be
 P(\rho) = a\rho^{\g}\quad (a>0,\g>1)
 \ee
$\mu$ and $\lambda$ are shear viscosity and bulk viscosity
respectively satisfying the physical condition:
\be
\mu>0, \lambda + \frac{2}{N}\mu\ge 0
\ee

   Lions \cite{L1} \cite{L2}, Feireisl \cite{F1}\cite{F2} et. established the global existence of
   weak solutions to the problem $(1.1)-(1.3)$, where vacuum is allowed initially.
    The global existence to the compressible Navier-Stokes equations is
   obtained by Matsumura\cite{M1}and Nishida under the condition that the initial data is a small perturbation
of a non-vacuum constant. It is also shown by Xin\cite{X1} that there is no global
in time regular solutions in $R^3$ to the compressible Naiver-Stokes equations provided the
initial density is compactly supported.

There are many results concerning the existence of strong solutions
to the Navier-Stokes equations, only local existence results have been established, see \cite{K1},
   \cite{K2},\cite{K3},\cite{S2}. V.A.Solonnikov proved in \cite{S1}
that for $C^2$ pressure laws and initial data satisfies for some
$q>N$, \be 0<m\le\rho_0(x)\le M<\infty,\quad and\quad \rho_0\in
W^{1,q}(T^N) \ee

\be u_0\in W^{2-\frac{2}{q},q}(T^N)^N \ee there exists a local unique
strong solution $(\rho,u)$ to $(1.5)-(1.6)$ for periodic data, such that
\be
\ba
& \rho\in L^{\infty}(0,T;W^{1,q}(T^N)),\quad \rho_t\in L^q((0,T)\times T^N)\\
& u\in L^q(0,T;W^{2,q}(T^N)),\quad u_t\in L^q((0,T)\times T^N)^N
\ea
\ee

Later, it was shown in \cite{K1} that if $\O$ is either a bounded
domain or the whole space, the initial data $\rho_0$
and $u_0$ satisfy \be 0\le\rho_0\in W^{1,\tilde{q}}(\O),\quad u_0\in
H_0^1(\O)\cap H^2(\O) \ee for some $\tilde{q}\in (3,\infty)$ and the
compatibility condition: \be -\mu\triangle u_0 - (\lambda +
\mu)\nabla\rm{div}u_0 + \nabla P(\rho_0) = \rho_0^{1/2}g \quad
for\quad some\quad g\in L^2(\O) \ee
then there exists a positive time $T_1\in (0,\infty)$ and a unique
strong solution $(\rho,u)$ to the isentropic problem, such that
\be
\ba
& \rho\in C([0,T_1];W^{1,q_0}(\O)),\quad \rho_t\in C([0,T_1];L^{q_0}(\O))\\
& u\in C([0,T_1];D_0^1\cap D^2(\O))\cap L^2(0,T_1;D^{2,q_0}(\O))\\
& u_t\in L^2(0,T_1;D_0^1(\O))\quad \sqrt{\rho}u_t\in L^{\infty}(0,T_1;L^2(\O))
\ea
\ee

 Furthermore, one has the
following blow-up criterion: if $T^*$ is the maximal time of
existence of the strong solution $(\rho,u)$ and $T^*<\infty$, then
\be \sup\limits_{t\rightarrow T^*}(\norm[W^{1,q_0}]{\rho} +
\norm[D_0^1]{u}) = \infty \ee where $q_0 = \min (6,\tilde{q})$.

Throughout this paper, we use the following notations for the standard
 homogeneous and inhomogeneous Sobolev spaces.

$$ D^{k,r}(\O) =\{u\in L^1_{loc}(\O):\norm[L^r]{\nabla^k u}<\infty\}, $$
$$W^{k,r} = L^r\cap D^{k,r}, H^k = W^{k,2},\quad D^k = D^{k,2}$$
$$D_0^1 = \{u\in L^6(\O):\norm[L^2]{\nabla u}<\infty\quad and\quad u=0\quad on\quad \p\O\},$$
$$ H_0^1 = L^2\cap D_0^1,\quad \norm[D^{k,r}]{u} = \norm[L^r]{\nabla^k u}$$

Very recently, Fan and Jiang \cite{J1} proved a blow-up
criterion for such strong solutions. i.e, when  $7\mu>9\lambda$,
\be
 \lim_{T\rightarrow T^*}\{\sup\limits_{0\le
T< T^*}\norm[L^{\infty}]{\rho} + \int_0^T(\norm[W^{1,q_0}]{\rho} +
\norm[L^2]{\nabla\rho}^4)dt\} = \infty
\ee
Here they only require a sufficient regularity of density $\rho$ to
 admit the global existence of strong solutions, as $(1.11)$ revealed.

In this paper, we assume that
\be
\ba
& \mu + \lambda\ge 0, \quad N=2, \quad \O = T^2 \\
& \mu + \lambda = 0, \quad N=3,\quad \O\subset R^3
\ea
\ee

Here and thereafter $C$ always denotes a generic constant depending only on
$\O,T$ and initial data.

For the initial boundary value problem, we have the following
result:

\begin{theorem}
Let $\O$ be either a three dimensional bounded domain or two dimensional torus.
 $Q_T = (0,T)\times\O$. Assume that the initial data
satisfy $(1.7)-(1.8)$ and $(1.12)$ holds. Let $(\rho,u)$ be a strong solution of
the problem $(1.1)-(1.3)$ satisfying the regularity $(1.9)$. If $T^*<\infty$ is the
maximal time of existence, then \be \lim_{T\rightarrow
T^*}\int_0^T\norm[L^{\infty}(\O)]{\nabla u}dt = \infty \ee
\end{theorem}

In the case of initial value problem, it holds that
\begin{theorem}
  Let $\O = R^3$. Assume that the initial data satisfy
\be
\rho_0\in H^1(R^3)\cap W^{1,\tilde{q}}(R^3),\quad u_0\in
D_0^1(R^3)\cap D^2(R^3)
\ee
 for some $\tilde{q}$ with $3<\tilde{q}<\infty$ and
the compatibility condition $(1.8)$. Let $(\rho,u)$ be a strong solution to the
problem $(1.1)-(1.3)$, and satisfy
\be
\ba
& \rho\in C([0,T_1],H^1(R^3)\cap W^{1,q_0}(R^3)),\quad \rho_t\in C([0,T_1],L^2(R^3)\cap L^{q_0}(R^3)) \\
& u\in C([0,T_1],D_0^1(R^3)\cap D^2(R^3))\cap L^2(0,T_1;D^{2,q_0}(R^3))\\
& u_t\in L^2(0,T_1;D_0^1),\quad \sqrt{\rho}u_t\in
L^{\infty}(0,T_1;L^2(R^3))
\ea
\ee
 where $q_0 = \min(6,\tilde{q})$.
If $T^*<\infty$ is the maximal time of existence, then \be
\lim_{T\rightarrow T^*}\int_0^T\norm[L^{\infty}(\O)]{\nabla u}dt =
\infty \ee
\end{theorem}

{\bf Remark 1.1} The blow up criterion $(1.10)$ involves both the density and velocity. It may be
natural to expect the higher regularity of velocity if density is regular enough. $(1.11)$ shows
that sufficient regularity of the gradient of density indeed guarantees the global existence of strong
solutions. The main difficulty in our case is to control the gradient of density, which is not a priori known
and coupled with the second derivative of velocity.

     We develop some new estimates under the condition $(2.1)$. In fact, the
key estimates in our analysis are both $L^{\infty}$ bound of $\rho u$ and $L^{\infty}(0,T;L^2(\O))$
norm of $\nabla\rho$. The Super-norm estimate for the momentum is obtained by a Moser-iteration
based on the a priori energy bounds motivated by an analysis in \cite{H1}. To control the
$L^{\infty}(0,T;L^2(\O))$ norm of $\nabla\rho$, our key observation is that the space-time square mean
of the convection term $F = \rho u_t + \rho u\cdot\nabla u$ is controlled by that of $\nabla\rho$
(see Lemma 2.3). This, in turn, gives the desired $L^{\infty}(0,T;L^2(\O))$ estimate on $\nabla\rho$,
and thus the $L^2(0,T;H^2(\O))$ of $u$. Then the higher order regularity can be obtained by using the
equations and the compatibility condition $(1.8)$.

{\bf Remark 1.2} There are many results concerning blow-up criteria of the incompressible flow.
In their well-known paper \cite{B1}, Beal-Kato-Majda established a blow-up
criterion for the incompressible Euler equations. One can get global smooth solution if
 $\int_0^T\norm[L^{\infty}]{\omega}dt$ is bounded. It's worth noting
 that only the vorticity $\omega$ plays an important role in the global
 existence of smooth solutions. Moreover, as pointed out by Constantin\cite{C5},
 the solution is smooth if and only if $\int_0^T\norm[L^{\infty}]{((\nabla u)\xi)\cdot \xi}$ is bounded,
 where $\xi$ is the unit vector in the direction of $\omega$. It turns out that the solution
 becomes smooth either the asymmetric
 or symmetric part of $\nabla u$ is controlled.
   Later, Constantin\cite{C3}, Fefferman and Majda
 showed a sufficient geometric condition to control the breakdown of smooth solutions of incompressible
 Euler involving the Lipschitz regularity of the direction of the vorticity. It is also shown by
 Constantin\cite{C4} and Fefferman that the solution of incompressible Navier-Stokes equations is smooth if
 the direction of vorticity is well behaved.

  Recently, in\cite{C1}, assuming that the added stress tensor is given in a proper form,
   and using an idea of J.-Y. Chemin and N. Masmoudi \cite{C2},
 Constantin, P. and Fefferman, C., Titi, E. S. and Zarnescu, A
 obtain a logarithmic bound for $\int_{0}^{T}\|\nabla u\|_{L^{\infty}}dt.$ to conclude
 that the solution to Navier-Stokes-Fokker-Planck system exists for all time and is smooth.

 In our paper, we establish a similar criterion to Beal-Kato-Majda. Our blow up criteria involve
 both the symmetric and asymmetric part of $\nabla u$, as compressibility and vorticity are
 two key issues in the formation of singularity of compressible Navier-Stokes.

{\bf Remark 1.3} The condition $(1.12)$ is assumed to obtain
 $L^{\infty}(0,T;L^{\infty}(\O))$ norm of the momentum $\rho u$ by a Moser-Iteration. However, in $2-D$
 periodic case, using an estimate motivated by Desjardin\cite{D1}, we can show $(1.13)$ holds for
 natural physical constraint: $\mu + \lambda\ge 0$.

{\bf Remark 1.4} We will study the blow up criteria for smooth solution of compressible Navier-Stokes
in another paper\cite{H2}.

\section{Proof of Theorem 1.1}

Let $(\rho,u)$ be a strong solution to the problem $(1.1)-(1.3)$. We
assume that the opposite  holds, i.e
\be \lim_{T\rightarrow
T^*}\int_0^T\norm[L^{\infty}(\O)]{\nabla u}dt\le C<\infty \ee
First, the standard energy estimate yields \be \sup\limits_{0\le t\le
T}\norm[L^2]{\rho^{1/2}u(t)} + \int_0^T\norm[H^1]{u}^2dt \le C,\quad 0\le
T<T^* \ee

By assumption $(2.1)$ and the conservation of mass, the $L^{\infty}$ bounds of density
follows immediately,
\begin{lemma}
  Assume that
\be
  \int_0^T\norm[L^{\infty}]{div u}dt\le C, \quad 0\le T<T^*
\ee

then  \be
  \norm[L^{\infty}(Q_T)]{\rho} \le C,\quad 0\le T<T^*
\ee
\end{lemma}
\proof It follows from the conservation of mass that for $\forall q>1$,
\be \p_t(\rho^q) +
\rm{div}(\rho^qu) + (q-1)\rho^q\rm{div}u = 0 \ee
Integrate $(2.5)$ over $\O$ to obtain,
\be \p_t\int_{\O}\rho^qdx  \le
(q-1)\norm[L^{\infty}(\O)]{\nabla u}\int_{\O}\rho^qdx \ee i.e \be
\p_t\norm[L^q]{\rho} \le \frac{q-1}{q}\norm[L^{\infty}(\O)]{\nabla
u}\norm[L^q]{\rho} \ee
 Which implies immediately
  \be
\norm[L^q]{\rho}(t)\le C \ee
with $C$ independent of $q$, so our lemma follows.

\endproof

In the next proposition, we derive bound on $L^{\infty}$ norm of momentum
$\rho u$.

\begin{proposition}
Under condition $(2.3)$, it holds that
\be
\norm[L^{\infty}(Q_T)]{\rho u}\le
C(\norm[L^{\infty}]{\rho_0},\norm[L^{\infty}]{u_0},
\norm[L^1L^{\infty}]{\nabla u}, T),\quad 0\le T<T^*
\ee
\end{proposition}

\proof Let $p$ be a fixed positive large number. Obviously,
\be
\int_{\O}\rho_0|u_0|^{p+2}dx\le c_0^{p+2}
\ee

Without losing of generality, we assume $\mu = 1$.

Multiplying $|u|^pu$ on both sides of the momentum equations in $(1.1)$ yields
\be
\ba & \frac{1}{p+2}\frac{d}{dt}\int_{\O}\rho |u|^{p+2}dx + \int_{\O}
|\nabla u|^2|u|^pdx + p\int_{\O} |u|^p(\nabla |u|)^2dx \\
& = \int_{\O}a\rho^{\g}|u|^pdiv u dx + p\int_{\O}a\rho^{\g}|u|^{p-1}u\cdot\nabla |u|dx\\
& = I + II \ea
\ee

First, it follows from $(2.4)$ and H\"{o}lder inequality that
\be
\ba |I| & \le C(\int_{\O}|u|^p|\nabla u|^2dx)^{1/2}(\int_{\O}\rho|u|^{p+2}dx)^{\frac{p}{2(p+2)}}\\
& \le \frac{1}{2}\int_{\O}|u|^p|\nabla u|^2dx +
C(\int_{\O}\rho|u|^{p+2}dx)^{\frac{p}{p+2}} \ea
\ee

Similarly,
\be
\ba |II| & \le Cp(\int_{\O}|u|^p|\nabla |u||^2dx)^{1/2}(\int_{\O}\rho|u|^{p+2}dx)^{\frac{p}{2(p+2)}}\\
& \le \frac{p}{2}\int_{\O}|u|^p(\nabla |u|)^2 +
Cp(\int_{\O}\rho|u|^{p+2}dx)^{\frac{p}{p+2}}  \ea
\ee

Therefore,
\be
\ba & \frac{1}{p+2}\p_t\int_{\O}\rho |u|^{p+2}dx +
\frac{1}{2}\int_{\O}|u|^p|\nabla u|^2dx
+ \frac{p}{2}\int_{\O}|u|^p(\nabla |u|)^2dx \\
& \le C(p+1)(\int_{\O}\rho|u|^{p+2}dx)^{\frac{p}{p+2}} \ea
\ee

Integrating $(2.14)$ over $(0,T)$ yields

\be
\ba & \frac{1}{p+2}\int_{\O}\rho |u|^{p+2}(T)dx +
\frac{1}{2}\int_{Q_T}|u|^p|\nabla u|^2dx
+ \frac{p}{2}\int_{Q_T}|u|^p(\nabla |u|)^2dx \\
& \le C(p+1)\int_0^T(\int_{\O}\rho|u|^{p+2}dx)^{\frac{p}{p+2}}dt +
\frac{1}{p+2}\int_{\O}
\rho_0|u_0|^{p+2}dx \\
& \le
C(p+1)(\int_{Q_T}\rho|u|^{p+2}dxdt)^{\frac{p}{p+2}}T^{\frac{2}{p+2}}
+ \frac{1}{p+2}\int_{\O}
\rho_0|u_0|^{p+2}dx \\
& \le C(p+1)(\int_{Q_T}\rho|u|^{p+2}dxdt)^{\frac{p}{p+2}} +
\frac{1}{p+2}c_0^{p+2} \ea
\ee
It follows that
$$
\int_{\O}\rho|u|^{p+2}(T)dx\le C(p,c_0,T)
$$

Moreover, it follows from H\"{o}lder inequality and $(2.15)$ that
\be
\ba \int_{Q_T}\rho|u|^{\frac{5}{3}(p+2)}dxdt & =
\int_{Q_T}(\rho^{\frac{3}{2}}|u|
^{p+2})^{\frac{2}{3}}|u|^{p+2}dxdt \\
& \le \int_0^T(\int_{\O}\rho^{\frac{3}{2}}|u|^{p+2}dx)^{\frac{2}{3}}
(\int_{\O}|u|^{3(p+2)}dx)^{\frac{1}{3}}dt \\
& \le C\int_0^T(\int_{\O}\rho|u|^{p+2}dx)^{\frac{2}{3}}(\int_{\O}|u|^{3(p+2)})^{\frac{1}{3}}dt\\
&\le C(\sup\limits_{0\le t\le
T}\int_{\O}\rho|u|^{p+2}dx)^{\frac{2}{3}}
\int_0^T(\int_{\O}(|u|^{\frac{p+2}{2}})^6dx)^{\frac{1}{3}}dt\\
&\le C(\sup\limits_{0\le t\le
T}\int_{\O}\rho|u|^{p+2}dx)^{\frac{2}{3}}
\int_0^T\int_{\O}(\nabla |u|^{\frac{p+2}{2}})^2dxdt\\
& \le (C(p+2)^2(\int_{Q_T}\rho|u|^{p+2})^{\frac{p}{p+2}} + c_0^{p+2})^{\frac{5}{3}}\\
& \le
C(p+2)^{\frac{10}{3}}(\int_{Q_T}\rho|u|^{p+2}dxdt)^{\frac{5}{3}}
  + C(p+2)^{\frac{10}{3}} + C_0^{\frac{5}{3}(p+2)}
\ea
\ee
Here $C_0 = 2c_0$. Set
\be
r=\frac{5}{3},p+2 = r^k,c_2=\frac{10}{3},
c_1=C(|\rho_0|_{L^{\infty}},
|u_0|_{L^{\infty}},\norm[L^1L^{\infty}]{\nabla u}, T)
\ee
 We conclude from above the following reverse Holder inequality.
\be
\int_{Q_T}\rho|u|^{r^{k+1}}dxdt \le
c_1r^{c_2k}(\int_{Q_T}\rho|u|^{r^k}dxdt)^r + c_1r^{c_2k} +
C_0^{r^{k+1}}
\ee

Define
\be
A(k) = \int_{Q_T}\rho|u|^{r^k}dxdt,\quad f(k) = c_1r^{c_2k},\quad g(k) = C_0^{r^k}
\ee
Then $(2.18)$ could be written as
\be
A(k+1)\le f(k)A(k)^r + f(k) + g(k)^r
\ee
Write
\be
B(k) = max(A(k),g(k),1),\quad F(k) = 3f(k)
\ee
Without lose of generality, we assume that $f(k)\ge 1$. Hence
\be
\ba
B(k+1) & = max(A(k+1),g(k+1),1)\\
& \le max(f(k)A(k)^r + f(k) + g(k)^r,g(k)^r,1)\\
& \le max(f(k)B(k)^r + f(k)B(k)^r + f(k)B(k)^r,g(k)^r,1)\\
& \le 3f(k)B(k)^r = F(k)B(k)^r
\ea
\ee
By induction, we obtain
\be
\ba
B(k+1) & \le F(k)B(k)^r\\
& \le F(k)(F(k-1)B(k-1)^r)^r = F(k)F(k-1)^rB(k-1)^{r^2}\\
& \le\ldots\ldots \\
& \le F(k)F(k-1)^rF(k-2)^{r^2}\ldots F(2)^{r^{k-2}}B(2)^{r^k}
\ea
\ee

Consequently,
\be
\ba
B(k+1)^{\frac{1}{r^{k+1}}} & \le F(k)^{\frac{1}{r^{k+1}}}F(k-1)^{\frac{1}{r^k}}\ldots
F(2)^{\frac{1}{r^3}}B(2)^{\frac{1}{r^2}}\\
& = (3c_1)^{\sum_{i=3}^{k+1}r^{-i}}r^{c_2\sum_{i=0}^{k-2}(k-i)r^{i-k-1}}B(2)^{\frac{1}{r^2}}
\ea
\ee
And
\be
\sum_{i=3}^{k+1}r^{-i} + \sum_{i=0}^{k-2}(k-i)r^{i-k-1} <\infty
\ee

\be
B(2) = max(A(2),g(2),1) = max((\int_{Q_T}\rho |u|^{r^2}dxdt)^{\frac{1}{r^2}},C_0^{r^2},1)<\infty
\ee
By definition
\be
(\int_{Q_T}\rho|u|^{r^k}dxdt)^{\frac{1}{r^k}}\le B(k)^{\frac{1}{r^k}}\le C
\ee

Thus we can conclude , $\forall q>1$, there exists a $k,\a$ with $0<\a\le1$, $r^{k-1}\le q<r^k$

$$\norm[L^q(Q_T)]{\rho u} \le \norm[L^{r^{k-1}}(Q_T)]{\rho u}^{\a}\norm[L^{r^k}(Q_T)]{\rho u}^{1-\a}$$

Note that
$$(\int_{Q_T}\rho^{r^k}|u|^{r^k}dxdt)^{\frac{1}{r^k}} \le \norm[L^{\infty}(Q_T)]{\rho}^{\frac{r^k-1}{r^k}}
(\int_{Q_T}\rho|u|^{r^k}dxdt)^{\frac{1}{r^k}}\le C$$

Consequently,
\be
\norm[L^{\infty}(Q_T)]{\rho u}\le C(|\rho_0|_{L^{\infty}},
|u_0|_{L^{\infty}},\norm[L^1L^{\infty}]{\nabla u},T)
\ee

Moreover, for $\alpha,\beta >0$, $p\ge 1$
\be
\norm[L^{\infty}(Q_T)]{\rho^{\alpha}|u|^{\beta}} \le C(\alpha,\beta),\quad
\norm[L^p(Q_T)]{\rho^{\a}|u|^{\beta}}\le C(\a,\beta,p)
\ee

\endproof

The next lemma shows a connection between a convection term and
the gradient of the density, which will play an important role in deriving the desired
bounds on $\nabla\rho$.
\begin{lemma}
  Let $F=\rho u_t + \rho u\cdot\nabla u$. Then it holds that
  $$\int_{Q_T}F^2dxdt \le C\int_{Q_T}|\nabla\rho|^2dxdt + C,\quad 0\le T<T^*$$
\end{lemma}

\proof

Note that
\be
\int_{Q_T}F^2dxdt \le C^*(\norm[L^{\infty}(Q_T)]{\rho})\int_{Q_T}\rho
u_t^2dxdt +
 2\int_{Q_T}|\rho u\cdot\nabla u|^2dxdt
\ee
It follows from proposition $2.1$ that
\be
\int_{Q_T}F^2dxdt \le C^*(\norm[L^{\infty}(Q_T)]{\rho})\int_{Q_T}\rho
u_t^2dxdt + C(\norm[L^{\infty}(Q_T)]{\rho u}^2)\int_{Q_T}|\nabla
u|^2dxdt
\ee

 Multiplying the momentum equation by $u_t$ and integrating show that
\be
\int_{\O}\rho u_t^2dx + \int_{\O}\rho u\cdot\nabla u\cdot u_tdx +
\frac{1}{2}\frac{d}{dt}\int_{\O}|\nabla u|^2dx = \int_{\O}P\rm{div}u_tdx
\ee

Note that
\be
\int_{\O}P\rm{div}u_tdx = \p_t\int_{\O}P\rm{div}udx -
\int_{\O}P_t\rm{div}udx,
\ee
and
$$P_t + \rm{div}(Pu) + (\g-1)P\rm{div}u = 0$$
One gets
\be
\ba \int_{\O}P\rm{div}u_tdx & = \p_t\int_{\O}P\rm{div}udx +
\int_{\O}\rm{div}(Pu)\rm{div}udx +
(\g-1)\int_{\O}P(\rm{div}u)^2dx\\
& = \p_t\int_{\O}P\rm{div}udx - \int_{\O}(Pu)\cdot\nabla\rm{div}udx
+ (\g-1)\int_{\O}P(\rm{div}u)^2dx \ea
\ee

Combing $(2.32)$ and $(2.34)$ yields
\be
\ba & \quad \frac{1}{2}\int_{\O}|\nabla u|^2dx(T) + \int_{Q_T}\rho
u_t^2dxdt
 + \int_{Q_T}\rho u\cdot\nabla u\cdot u_tdxdt  \\
& = \frac{1}{2}\int_{\O}|\nabla u_0|^2dx + \int_{\O}P\rm{div}udx(T)
-
\int_{\O}P_0\rm{div}u_0dx\\
& - \int_{Q_T}Pu\cdot\nabla\rm{div}udxdt +
(\g-1)\int_{Q_T}P(\rm{div}u)^2dxdt \ea
\ee

Direct estimates show that
\be
\int_{\O}P\rm{div}udx(T)\le \frac{1}{4}\int_{\O}|\nabla u|^2dx(T) +
C
\ee

\be
\ba \int_{Q_T}\rho u\cdot\nabla u\cdot u_tdxdt & \le
\frac{1}{2}\int_{Q_T}\rho u_t^2 +
\int_{Q_T}\rho |u\cdot\nabla u|^2dxdt\\
&\le \frac{1}{2}\int_{Q_T}\rho u_t^2 + C(\norm[L^{\infty}(Q_T)]{\rho|u|^2})\int_{Q_T}|\nabla u|^2dxdt \\
& = \frac{1}{2}\int_{Q_T}\rho u_t^2 + C \ea
\ee

On the other hand, it follows from $F = \lap u - \nabla P$ and $(2.2)$ that
\be
\ba \int_{Q_T}Pu\cdot\nabla\rm{div}udxdt & =
\int_{Q_T}Pu\cdot\nabla\rm{div}\triangle^{-1}
\nabla Pdxdt + \int_{Q_T}Pu\cdot\nabla\rm{div}\triangle^{-1}Fdxdt\\
& \le C\int_{Q_T}|\nabla\rho|^2dxdt + \ep\int_{Q_T}F^2dxdt + C \ea
\ee

Consequently,
\be
\int_{Q_T}\rho u_t^2dxdt + \frac{1}{2}\int_{\O}|\nabla u|^2dx(T) \le
C\int_{Q_T}|\nabla\rho|^2dxdt + 2\ep\int_{Q_T}F^2dxdt + C
\ee

Choosing $\ep$ as $2C^*\ep <1$, we may conclude
$$\int_{Q_T}F^2dxdt \le C\int_{Q_T}|\nabla\rho|^2dxdt + C$$
Which completes the proof of Lemma $2.3$.

\endproof
We are now ready to show the desired $L^{\infty}(0,T;L^2(\O))$ estimate of $\nabla\rho$.
\begin{proposition}
 Under the assumption $(2.1)$, it holds that
  \be
  \sup\limits_{0\le t\le T}\int_{\O}|\nabla\rho|^2dx \le C \quad, 0\le T<T^*
  \ee
\be
\int_{Q_T}\rho u_t^2dxdt + \sup\limits_{0\le t\le T}\int_{\O}|\nabla
u|^2dx\le C \quad, 0\le T<T^*
\ee
\be
\int_0^T\norm[H^2(\O)]{u}^2dt\le C,\quad 0\le T<T^*
\ee
\end{proposition}

\proof Differentiating the mass equation in $(1.1)$ with respect to $x_i$,
$$\p_t(\p_i\rho) + \rm{div}(\p_i\rho u) + \rm{div}(\rho\p_i u) = 0$$
Which can be multiplied by $2\p_i\rho$ to obtain
\be
\p_t|\p_i\rho|^2 + \rm{div}(|\p_i\rho|^2u) + |\p_i\rho|^2\rm{div}u +
2\p_i\rho\rho\p_i\rm{div}u + 2\p_i\rho\p_i u\cdot\nabla\rho = 0
\ee
Integrating over $\O$ and using $\rm{div}u =
\rm{div}\triangle^{-1}\nabla P + \rm{div}\triangle^{-1}F$ show that

\be
\ba \p_t\int_{\O}|\p_i\rho|^2dx & =
-\int_{\O}|\p_i\rho|^2\rm{div}udx - 2\int_{\O}\rho\p_i\rho
\p_i\rm{div}\triangle^{-1}\nabla Pdx\\
& - \int_{\O}2\rho\p_i\rho\p_i\rm{div}\triangle^{-1}Fdx - \int_{\O}2\p_i\rho\p_iu\cdot\nabla\rho dx\\
& = -(A_1 + A_2 + A_3 + A_4) \ea
\ee

Each term on the right hand side of $(2.44)$ can be estimated as follows:

\be
|A_1(t)|\le \norm[L^{\infty}]{div u}(t)\int_{\O}|\p_i\rho|^2dx
\le\norm[L^{\infty}]{div u}(t)\int_{\O}|\nabla\rho|^2dx
\ee

\be
|A_2(t)|\le C\norm[L^2]{\nabla\rho}\norm[L^2]{\nabla P}\le
C\int_{\O}|\nabla\rho|^2dx
\ee

\be
|A_3(t)|\le C\norm[L^2]{\nabla\rho}\norm[L^2]{F}
\ee

\be
|A_4(t)|\le C\norm[L^{\infty}]{\nabla u}(t)\int_{\O}|\nabla\rho|^2dx
\ee

Consequently,
\be
\p_t\int_{\O}|\nabla\rho|^2dx \le C(\norm[L^{\infty}]{\nabla u}(t) +
1) \int_{\O}|\nabla\rho|^2dx + C\int_{\O}F^2dx
\ee

This, together with Gronwall's inequality yields
\be
\ba \int_{\O}|\nabla\rho|^2dx(t) & \le
Ce^{C\int_0^t(\norm[L^{\infty}]{\nabla u}(s) + 1)ds}
(\int_{\O}|\nabla\rho_0|^2dx + \int_0^t(\int_{\O}F^2(s)dx)
e^{-C\int_0^s(\norm[L^{\infty}]{\nabla u}(\tau) + 1)d\tau}ds)\\
&\le C\int_0^t\int_{\O}F^2dxds + C\\
&\le C\int_0^t\int_{\O}|\nabla\rho|^2dxds + C \ea
\ee

Hence
\be
\sup\limits_{0\le t\le T}\int_{\O}|\nabla\rho|^2dx \le C
\ee

Next, it follows from $(2.31),(2.39)$ and $(2.51)$ that
\be
\int_{Q_T}\rho u_t^2dxdt + \sup\limits_{0\le t\le T}\int_{\O}|\nabla
u|^2dx\le C
\ee
This, together with $\triangle u = \rho u_t + \rho u\cdot\nabla u +
\nabla P$, shows that

\be
\ba \norm[L^2(0,T;H^2(\O))]{u} & \le \norm[L^2(Q_T)]{\rho u_t} +
\norm[L^2(Q_T)]{\rho u\cdot\nabla u} +
\norm[L^2(Q_T)]{\nabla P} \\
& \le C + C\norm[L^2(Q_T)]{\nabla u} +
C\norm[L^2(Q_T)]{\nabla\rho}\le C \ea
\ee

\endproof

Next, we proceed to improve the regularity class of $\rho$ and
$u$. To this end, we first derive some bounds on derivatives of $u$ based on
above estimates.

\begin{proposition}
 Under the condition $(2.1)$, it holds that
  \be
  \sup\limits_{0\le t\le T}\norm[L^2]{\rho^{1/2}u_t(t)}^2 + \int_{Q_T}|\nabla u_t|^2dxdt \le C,
  \quad 0\le T<T^*
  \ee
\be
  \sup\limits_{0\le t\le T}\norm[H^2]{u}\le C,\quad 0\le T<T^*
\ee
\end{proposition}

\proof  Differentiating the momentum equations in $(1.1)$ with respect to time $t$ yields
\be
\rho u_{tt} + \rho u\cdot\nabla u_t - \triangle u_t + \nabla p_t =
-\rho_t(u_t + u\cdot\nabla u) - \rho u_t\cdot\nabla u
\ee
 Taking the inner product of the above equation with $u_t$ in $L^2(\O)$ and integrating by parts, one gets
\be
\ba
& \frac{d}{dt}\int_{\O}\frac{1}{2}\rho u_t^2dx + \int_{\O}|\nabla u_t|^2dx - \int_{\O}P_t\rm{div}u_tdx \\
& = -\int_{\O}(\rho u\cdot\nabla[(u_t + u\cdot\nabla u)u_t] +
\rho(u_t\cdot\nabla u)\cdot u_t)dx \ea
\ee

The last term on the left-hand side of $(2.57)$ can be rewritten as (using $(2.33)$):
\be
\ba -\int_{\O}P_t\rm{div}u_tdx & =
\frac{d}{dt}\int_{\O}\frac{\g}{2}P(\rm{div}u)^2dx +
\int_{\O}\nabla P\cdot(u\rm{div}u_t)dx \\
& + \frac{\g}{2}\int_{\O}(-Pu\cdot\nabla(\rm{div}u)^2 +
(\g-1)P(\rm{div}u)^3)dx \ea
\ee

It follows from $(2.57)$ and $(2.58)$ that
\be
\ba & \frac{d}{dt}\int_{\O}(\frac{1}{2}\rho u_t^2 +
\frac{\g}{2}P(\rm{div}u)^2)dx +
\int_{\O}|\nabla u_t|^2dx\\
& \le \int_{\O}(2\rho|u||u_t||\nabla u_t| + \rho|u||u_t||\nabla u|^2
+ \rho|u|^2|u_t||\nabla^2u|
+ \rho|u|^2||\nabla u||\nabla u_t| \\
& + \rho|u_t|^2|\nabla u| + |\nabla P||u||\nabla u_t| + \g P|u||\nabla u||\nabla^2u| + \g^2P|\nabla u|^3)dx\\
& \equiv \sum_{i=0}^{8}F_i \ea
\ee

Now, we estimate each $F_i$ separately, where the Sobolev inequality and H\"{o}lder inequality will
be frequently used.

\be
\ba
|F_1| & = \int_{\O}2\rho|u||u_t||\nabla u_t|dx\\
& \le C\norm[L^{\infty}(Q_T)]{\rho^{1/2} u}\norm[L^2]{\rho^{1/2}u_t}\norm[L^2]{\nabla u_t}\\
& \le \ep\norm[L^2]{\nabla u_t}^2 + C\norm[L^2]{\rho^{1/2}u_t}^2 \ea
\ee

Due to $(2.40)$ and $(2.41)$, one has
 $$\sup\limits_{0\le t\le T}\int_{\O}|\nabla u|^2dx \le C$$
Thus, it follows from H\"{o}lder inequality, Sobolev imbedding and interpolation inequality that
\be
\ba
|F_2| & = \int_{\O}\rho|u||u_t||\nabla u|^2dx \\
& \le C\norm[L^{\infty}(Q_T)]{\rho^{1/2} u}\int_{\O}\rho^{1/2}|u_t||\nabla u|^2dx\\
& \le C\norm[L^3]{\rho^{1/2}u_t}\norm[L^3]{\nabla u}^2\\
& \le C\norm[L^2]{\rho^{1/2}u_t}^{1/2}\norm[L^6]{\rho^{1/2}u_t}^{1/2}
\norm[L^2]{\nabla u}\norm[L^6]{\nabla u}\\
& \le \ep\norm[L^2]{\rho^{1/2}u_t}\norm[L^2]{\nabla u_t} + C\norm[L^6]{\nabla u}^2\\
& \le \ep\norm[L^2]{\nabla u_t}^2 + C\norm[L^2]{\rho^{1/2}u_t}^2 +
C\norm[L^2]{\nabla u}^2
+ C\norm[L^2]{\nabla^2 u}^2\\
& \le \ep\norm[L^2]{\nabla u_t}^2 + C\norm[L^2]{\rho^{1/2}u_t}^2 +
C\norm[H^2]{u}^2, \ea
\ee

\be
\ba
|F_3| & = \int_{\O}\rho|u|^2|u_t||\nabla^2 u|dx\\
& \le C\norm[L^2]{\rho^{1/2} u_t}\norm[L^2]{\nabla^2u}\\
& \le C\norm[L^2]{\rho^{1/2} u_t}^2 + C\norm[L^2]{\nabla^2u}^2, \ea
\ee

\be
\ba
|F_4| & = \int_{\O}\rho|u|^2|\nabla u||\nabla u_t|dx \\
& \le \ep\norm[L^2]{\nabla u_t}^2 + C\norm[L^2]{\nabla u}^2, \ea
\ee

\be
\ba
|F_5| & = \int_{\O}\rho|u_t|^2|\nabla u|dx \\
& \le C\norm[L^2]{\rho u_t^2}\norm[L^2]{\nabla u}\\
& \le C\norm[L^4]{\rho^{1/2}u_t}^2\\
& \le \ep\norm[L^6]{\rho^{1/2}u_t}^2 + C\norm[L^2]{\rho^{1/2} u_t}^2\\
& \le \ep\norm[L^6]{u_t}^2 + C\norm[L^2]{\rho^{1/2} u_t}^2 \\
& \le \ep\norm[L^2]{\nabla u_t}^2 + C\norm[L^2]{\rho^{1/2} u_t}^2,
\ea
\ee

\be
\ba
|F_6| & = \int_{\O}|\nabla P||u||\nabla u_t|dx\\
& \le C\int_{\O}|\nabla\rho||\nabla u_t|dx\\
& \le \ep\norm[L^2]{\nabla u_t}^2 + C\norm[L^2]{\nabla\rho}^2, \ea
\ee

\be
\ba
|F_7| & = \int_{\O}\g P|u||\nabla u||\nabla^2 u|dx \\
& \le C\norm[L^2]{\nabla u}\norm[L^2]{\nabla^2u}, \ea
\ee

Finally, noting that $\nabla u\in L^{\infty}(0,T;L^2(\O))\cap
L^1(0,T;L^{\infty}(\O))$,
$$L^3(Q_T)\subset L^{\infty}(0,T;L^2(\O))\cap L^1(0,T;L^{\infty}(\O))$$
Hence
\be
\ba
|F_8| & = \int_{\O}\g^2P|\nabla u|^3dx\\
& \le C\int_{\O}|\nabla u|^3dx \\
& \le C\norm[L^{\infty}(\O)]{\nabla u}\int_{\O}|\nabla u|^2dx\\
& \le C\norm[L^{\infty}(\O)]{\nabla u}, \ea
\ee

Collecting all the estimates for $F_i$, we conclude
\be
\ba & \frac{d}{dt}\int_{\O}(\frac{1}{2}\rho u_t^2 +
\frac{\g}{2}p(\rm{div}u)^2)dx +
\int_{\O}|\nabla u_t|^2dx\\
& \le 5\ep\int_{\O}|\nabla u_t|^2dx + C(\norm[L^2]{\rho^{1/2}u_t}^2
+ \norm[H^2]{u}^2 + \norm[L^2]{\nabla\rho}^2 +
\norm[L^{\infty}]{\nabla u}) \ea
\ee

Thanks to the compatibility condition:
\be
\rho_0(x)^{\frac{1}{2}}(\rho_0(x)^{\frac{1}{2}}u_t(t=0,x)
+ \rho_0^{\frac{1}{2}}u_0\cdot\nabla u_0(x) - \rho_0^{\frac{1}{2}}g) = 0
\ee
it holds that
\be
\rho_0(x)^{\frac{1}{2}}u_t(t=0,x) =
 \rho_0^{\frac{1}{2}}u_0\cdot\nabla u_0(x) - \rho_0^{\frac{1}{2}}g \in L^2(\O)
\ee

Therefore, for arbitrary small $\ep$, $(2.68)$ yields
\be
\sup\limits_{0\le t\le T}\norm[L^2]{\rho^{1/2}u_t(t)}^2 + \int_{Q_T}|\nabla u_t|^2dxdt \le C,
  \quad 0\le T<T^*
\ee
Moreover,
$$
\triangle u = \rho u_t + \rho u\cdot\nabla u + \nabla P\in L^{\infty}L^2
$$
Hence,
\be
\sup\limits_{0\le T< T^*}\norm[H^2]{u}^2\le C
\ee
Our lemma follows immediately.

\endproof

Finally, the following lemma gives bounds of derivatives of the
density and the second derivatives of the velocity.

\begin{lemma}
Under the condition $(2.1)$, it holds that
$$\sup\limits_{0\le t\le T}(\norm[L^{q_0}]{\rho_t(t)} + \norm[W^{1,q_0}]{\rho})\le C,\quad 0\le T<T^*$$
$$\int_0^T\norm[W^{2,q_0}]{u(t)}^2dt\le C,\quad 0\le T<T^*, q_0 = min(6,\tilde{q})$$
\end{lemma}

\proof It follows from $(2.71)$ and $(2.72)$ that
$$u_t\in L^2(0,T;L^6(\O)), \nabla u\in L^6(Q_T)$$
$$F\in L^2(0,T;L^6(\O))$$

Differentiating the mass equation in $(1.1)$ with
respect to $x_i$, one gets
$$\p_t(\p_i\rho) + \rm{div}(\p_i\rho u) + \rm{div}(\rho\p_i u) = 0$$
Multiplying above identity by $q_0|\p_i\rho|^{q_0-2}\p_i\rho$,
one gets that for some $q_0 = min(6,\tilde{q})$,
\be
\ba
& \p_t|\p_i\rho|^{q_0} + \rm{div}(|\p_i\rho|^{q_0}u) +
(q_0-1)|\p_i\rho|^{q_0}\rm{div}u \\
& + q_0|\p_i\rho|^{q_0-2}\p_i\rho\rho\p_i\rm{div}u +
q_0|\p_i\rho|^{q_0-2}\p_i\rho\p_i u\cdot\nabla\rho = 0
\ea
\ee
This, together with $\rm{div}u =
\rm{div}\triangle^{-1}\nabla P + \rm{div}\triangle^{-1}F$ shows that

\be
\ba \p_t\int_{\O}|\p_i\rho|^{q_0}dx & =
-(q_0-1)\int_{\O}|\p_i\rho|^{q_0}\rm{div}udx -
q_0\int_{\O}\rho|\p_i\rho|^{q_0-2}\p_i\rho\p_i\rm{div}\triangle^{-1}\nabla Pdx\\
& -
\int_{\O}q_0\rho|\p_i\rho|^{q_0-2}\p_i\rho\p_i\rm{div}\triangle^{-1}Fdx
-
q_0\int_{\O}|\p_i\rho|^{q_0-2}\p_i\rho\p_iu\cdot\nabla\rho dx\\
& = -(B_1 + B_2 + B_3 + B_4) \ea
\ee

Which can be estimated as:

\be
|B_1(t)|\le (q_0-1)\norm[L^{\infty}]{\nabla
u}(t)\int_{\O}|\p_i\rho|^{q_0}dx \le C\norm[L^{\infty}]{\nabla
u}(t)\int_{\O}|\nabla\rho|^{q_0}dx
\ee

\be
|B_2(t)|\le
C\norm[L^{\frac{q_0}{q_0-1}}]{|\nabla\rho|^{q_0-1}}\norm[L^{q_0}]{\nabla P}\le
C\int_{\O}|\nabla\rho|^{q_0}dx
\ee

\be
\ba
& |B_3(t)|\le C\norm[L^{\frac{q_0}{q_0-1}}]{|\nabla\rho|^{q_0-1}}\norm[L^{q_0}]{F}\\
& \le C\norm[L^{q_0}]{\nabla\rho}^{q_0-1}\norm[L^{q_0}]{F} \ea
\ee

\be
|B_4(t)|\le C\norm[L^{\infty}]{\nabla u}(t)\int_{\O}|\nabla\rho|^{q_0}dx
\ee

It follows from $(2.74)-(2.78)$ that
\be
\p_t\norm[L^{q_0}]{\nabla\rho} \le C(\norm[L^{\infty}]{\nabla u}(t) +
1)\norm[L^{q_0}]{\nabla\rho} + C\norm[L^{q_0}]{F}
\ee

Hence,
$$\sup\limits_{0\le t\le T}\norm[L^{q_0}]{\nabla\rho}\le C$$
Therefore, due to this, $(2.71)$ and interpolation inequality, one has
\be
\rho_t = -(u\cdot\nabla\rho + \rho\rm{div}u)\in L^{\infty}L^{q_0}
\ee
Recall that
$$
\triangle u = F +\nabla P\in L^2L^{q_0}
$$
gives
\be
\int_0^T\norm[W^{2,q_0}(\O)]{u}^2dt \le C
\ee
This will close the estimates and guarantee to have an
extension of the strong solution.

In fact, in view of Proposition $2.4 - 2.5$ and lemma $2.6$, the functions
$(\rho,u)|_{t = T^*} = \lim_{t\rightarrow T^*}(\rho,u)$ satisfy the conditions
imposed on the initial data $(1.7)-(1.8)$ at the time $t=T^*$ Furthermore,
$$
\rho^{\frac{1}{2}}u_t + \rho^{\frac{1}{2}}u\cdot\nabla u \in L^{\infty}L^2
$$
\be
-\lap u + \nabla P|_{t = T^*} = \lim_{t\rightarrow T^*}\rho^{\frac{1}{2}}
(\rho^{\frac{1}{2}}u_t + \rho^{\frac{1}{2}}u\cdot\nabla u)
\triangleq \rho^{\frac{1}{2}}g|_{t = T^*}
\ee
Where $g|_{t = T^*}\in L^2(\O)$. Therefore, we can take $(\rho,u)|_{t = T^*}$
as the initial data and apply the local existence theorem \cite{K1} to extend
our local strong solution beyond $T^*$. This contradicts the assumption on $T^*$.

\endproof

Note that after some minor modifications, the above ideas also works in both
periodic case and $\O=R^3$, so theorem $1.2$ holds.

When $\O = T^2$, using an estimate from Desjardin\cite{D1},
\be
\ba
& \int_0^T(\norm[L^2(T^2)]{\rho^{\frac{1}{2}}u_t}^2 +
\norm[L^2(T^2)]{\rho^{\frac{1}{2}}u\cdot\nabla u}^2)ds + \sup_{0\le t\le T}\norm[L^2(T^2)]{\nabla u(t,)}^2\\
& \le \exp(C\exp(\int_0^T\norm[L^{\infty}(T^2)]{\rho}\delta(s)ds))
\ea
\ee

Where
\be
\delta(s) = 1+ \norm[L^2(T^2)]{\nabla u}\in L^1(0,T)
\ee

it follows from $(2.83)-(2.84)$ that
\be
\int_{Q_T}F^2dxdt + \sup_{0\le t\le T}\norm[L^2(T^2)]{\nabla u}^2 \le C
\ee
Then we can do a similar estimate step by step as three dimensional case to
 obtain a higher regularity of $(\rho,u)$. We omit the detail for simplicity.

{\bf Acknowledgement}  This research is supported in part by Zheng Ge Ru Foundation,
Hong Kong RGC Earmarked Research Grants CUHK4028/04P, CUHK4040/06P, CUHK4042/08P, and
the RGC Central Allocation Grant CA05/06.SC01.

\end{document}